 \documentclass[5p,twocolumn]{elsarticle}

\usepackage{amssymb}

\journal{Nuclear Instruments and Methods A}

\begin{document}

\begin{frontmatter}



\title{Hydrophobic silica aerogel production at KEK}


\author[First,Second]{Makoto Tabata\corref{cor1}}
\ead{makoto@hepburn.s.chiba-u.ac.jp}
\cortext[cor1]{Corresponding author.} 
\author[Third]{Ichiro Adachi}
\author[Second]{Hideyuki Kawai}
\author[Fourth]{Takayuki Sumiyoshi}
\author[Fifth]{Hiroshi Yokogawa}

\address[First]{Institute of Space and Astronautical Science (ISAS), Japan Aerospace Exploration Agency (JAXA), Sagamihara, Japan}
\address[Second]{Department of Physics, Chiba University, Chiba, Japan}
\address[Third]{Institute of Particle and Nuclear Studies (IPNS), High Energy Accelerator Research Organization (KEK), Tsukuba, Japan}
\address[Fourth]{Department of Physics, Tokyo Metropolitan University, Hachioji, Japan}
\address[Fifth]{Advanced Materials Development Department, Panasonic Electric Works Co.,Ltd., Kadoma, Japan}

\begin{abstract}
We present herein a characterization of a standard method used at the High Energy Accelerator Research Organization (KEK) to produce hydrophobic silica aerogels and expand this method to obtain a wide range of refractive index ($n$ = 1.006$–-$1.14). We describe in detail the entire production process and explain the methods used to measure the characteristic parameters of aerogels, namely the refractive index, transmittance, and density. We use a small-angle X-ray scattering (SAXS) technique to relate the transparency to the fine structure of aerogels.
\end{abstract}

\begin{keyword}
Silica aerogel \sep Refractive index \sep Cherenkov radiator \sep Dust collector \sep SAXS
\end{keyword}

\end{frontmatter}


\section{Introduction}
\label{}
Silica aerogel is a useful solid material used in various scientific instruments because of its properties such as transparency, low bulk density, and unique refractive index. The mechanical and optical characteristics originate from an amorphous three-dimensional structure of silica particles and pores. The bulk density of silica aerogel is determined by the ratio of silica and pore air and can be freely adjusted in the production process. The porosity reaches up to 99.7\% or more. The refractive index $n$ of silica aerogel is proportional to its density $\rho $: $n-1 = k\rho $, where $k$ is a constant.

As reported by a large number of articles in this journal, one of the most successful applications of silica aerogel is as a Cherenkov radiator. The first use of aerogels as a Cherenkov medium was reported in 1974 \cite{cite1}. At that time, it was mentioned that the lowest refractive index available for aerogels was $n = 1.01$ and the highest was $n = 1.20$, which was attained by heating. In addition, ultralow-density aerogels were fabricated at the Jet Propulsion Laboratory, California Institute of Technology for collecting comet dust, and the Stardust spacecraft by NASA successfully retrieved dust samples from the comet Wild 2 in 2006 \cite{cite2}. Recently, aerogels as dust capturers have also been applied in fusion plasmas \cite{cite3}. Furthermore, silica aerogel is considered to be a muonium-emitting material and is used as a source of ultracold muon beams \cite{cite4}. The need for silica aerogel is thus increasing.

The history of aerogel development in Japan dates back to the 1980s \cite{cite5}. At present, the study of aerogel production in Japan is carried forward by a collaboration involving Japan's High Energy Accelerator Research Organization (KEK), Chiba University, and Panasonic (Matsushita) Electric Works. As described in this article, we can manufacture silica aerogel with $n$ = 1.006$–-$1.14 directly by our conventional production method, without heating. Our aerogels are installed not only in compact trigger-veto counters but also in large systems for charged-hadron identification as Cherenkov radiators in high-energy, nuclear, and cosmic-ray experiments all over the world. For example, aerogels with $n = 1.03$ were used for an electron-positron veto-threshold Cherenkov counter in the Laser Electron Photon Experiment at SPring-8 (LEPS) \cite{cite6} and for the detection of cosmic-ray antiprotons in the Balloon-borne Experiment with Superconducting Spectrometer (BESS) \cite{cite7}. Also, Matsushita $n = 1.03$ aerogel was employed for a ring imaging Cherenkov (RICH) counter to separate pions and kaons in the HERMES experiment at DESY \cite{cite8}. Moreover, our low-density aerogels were used as the space debris capture medium in a series of Micro-Particles Capturer (MPAC) experiments by JAXA (NASDA) on the International Space Station (ISS) \cite{cite9}.

As already reported in our previous articles \cite{cite10,cite11,cite12}, we are using new production methods to strongly push forward the development of aerogels with ultrahigh and ultralow densities. The novel aerogels will be used in Belle II, which is a super $B$-factory experiment at KEK \cite{cite13}, and for the Tanpopo, which is an astrobiological mission planned for the Japanese Experiment Module (JEM) in the ISS \cite{cite14}, in addition to the next generation of nuclear experiments at J-PARC. The new production methods will further widen the opportunities to employ aerogels in scientific instruments. However, our conventional production method remains a key technique to produce high-quality aerogels for various applications, and provides the basis for new production methods. Here we will review the conventional production method and its capabilities.

In this article, we summarize the conventional and standard technique for producing hydrophobic silica aerogel at KEK and clarify the key features of this technique. The methods we use to make basic optical measurements are also described. The optical properties (i.e., refractive index and transmittance) are given as important parameters characterizing aerogels. Finally, we discuss the relationship between the optical properties and the fine structure of aerogels based on bulk density measurements and a small-angle X-ray scattering (SAXS) technique.

All aerogel samples described in this article were manufactured by one of the authors (M.T.) at Chiba University and examined at KEK between 2004 and 2011. Typical dimensions of aerogel tiles were 10 $\times $ 10 $\times $ 2 cm$^3$ and the smallest sample from trial production had dimensions of 4 $\times $ 4 $\times $ 1 cm$^3$. Our production method allows us to produce large monolithic tiles. For example, 26 $\times $ 18 $\times $ 2 cm$^3$ aerogel tiles with $n \sim $ 1.05 were successfully manufactured in collaboration with Panasonic Electric Works in 2009. In 2004, chemically combined multilayer aerogel tiles with different refractive indices were tested at $n < 1.01$ to produce ultralow-density aerogels \cite{cite10}. The technique was immediately applied in a range of $1.045 < n < 1.055$ \cite{cite15} based on the concept of a proximity-focusing RICH counter with a multiple-refractive-index aerogel radiator \cite{cite16}. Thus, we currently retain the production technique and facility for producing various aerogels with a wide range of refractive indices.

\section{Silica-aerogel production method}
\label{}
Our present method (the third method, following the single- and two-step methods) of producing silica aerogel, developed in the 1990s, is based on the KEK method \cite{cite17}. To construct threshold aerogel Cherenkov counters (ACCs) \cite{cite18} for the Belle detector \cite{cite19}, hydrophobic aerogel with $n$ = 1.01$-–$1.03 was developed based on the KEK method. A total of 2 m$^3$ of aerogel was installed in the ACC in 1998 and, until the end of the Belle experiment in 2010, the ACC system played an important role in separating pions from kaons. In 2004, we modernized the KEK method by introducing the solvent \textit{N},\textit{N}-dimethylformamide [DMF, $HCON(CH_3)_2$] in the wet-gel (solvogel) synthesis process \cite{cite15}. With the modernized KEK method, we can produce highly transparent aerogels with $n \sim 1.05$ \cite{cite20}, and we have since been working to optimize chemical-preparation recipes. The modernized KEK method has four important features:
\begin{itemize}
 \item Simple solvogel synthesis by methyl silicate 51;
 \item Selection of solvents for solvogel synthesis according to the desired refractive index;
 \item Hydrophobic treatment by hexamethyldisilazane;
 \item Supercritical drying by carbon dioxide.
\end{itemize}

The fundamental process of solvogel synthesis is based on the following two successive reactions:
\[ Si(OCH_3)_4 + 4H_2O \to Si(OH)_4 + 4CH_3OH, \]
\[ \textit{m}Si(OH)_4 \to (SiO_2)_\textit{m} + 2\textit{m}H_2O. \]
Tetramethoxysilane [$Si(OCH_3)_4$] can be simultaneously hydrolyzed, condensed, and polymerized in appropriate solvents to create silica solvogel with the help of a basic catalyst (ammonia aqueous solution). Instead of tetramethoxysilane, we use the commercially available "methyl silicate 51" (MS51) to simplify the solvogel synthesis process. MS51 is prepared by polymerizing tetramethoxysilane into an oligomer (average degree of polymerization is 4), which has a high silica content (51\% by weight).

The selection of solvents, ethanol, methanol, and DMF for the solvogel synthesis affects the optical performance of aerogels. The relative quantity of solvents to total solvogel volume is the principle factor determining the aerogel refractive index. For high transparency aerogels, appropriate solvents should be selected according to the desired refractive index. As expected from the chemical equation above, the most basic solvent is methanol, which is used in a wide range of $n \ge 1.02$. However, at $n \sim 1.02$, shrinkage of aerogels synthesized with methanol increases and their transparency decreases. This problem was solved by using ethanol (99.5) for $n < 1.02$ \cite{cite17}. On the other hand, transparency in the high-refractive-index range was improved by introducing DMF, which was used mixed with methanol for $n < 1.06$ and alone for $n \ge 1.06$ \cite{cite15}. So far, DMF has been introduced in the range of $n \ge 1.03$.

\begin{figure}[t] 
\centering 
\includegraphics[width=0.50\textwidth,keepaspectratio]{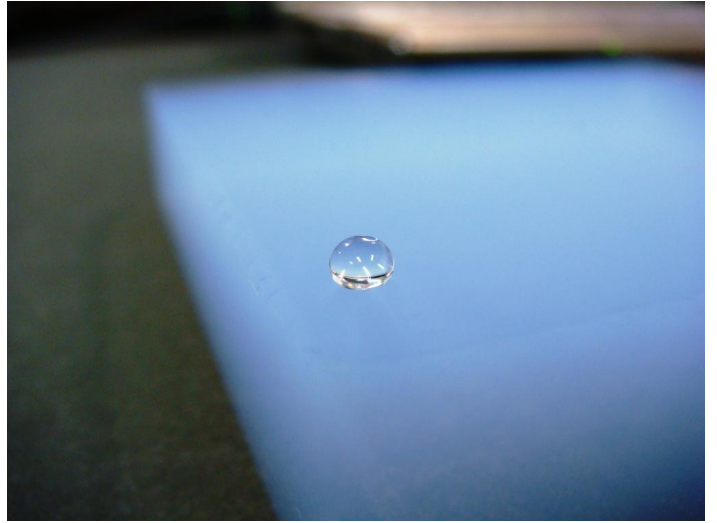}
\caption{Water drop on hydrophobic (water-resistant) aerogel (ID = LGG3-4b, $n = 1.009$).}
\label{fig:fig1}
\end{figure}

\begin{table*}[t]
\centering 
\caption{Starting materials used in aerogel production process.}
\label{table:table1}
	\begin{tabular}{ll}
		\hline
		Materials & Manufacturer (Distributor) \\
		\hline
		Methyl silicate 51 & Fuso Chemical Co., Ltd. \\
		Distilled water & Wako Pure Chemical Industries, Ltd. \\
		Ethanol (99.5)* & Wako Pure Chemical Industries, Ltd. \\
		Methanol & Wako Pure Chemical Industries, Ltd. \\
		\textit{N},\textit{N}-dimethylformamide & Wako Pure Chemical Industries, Ltd. \\
		28\% Ammonia solution & Wako Pure Chemical Industries, Ltd. \\
		Silazane, Z-6079 & Dow Corning Toray Co., Ltd. \\
		Ethanol (99)** & Japan Alcohol Trading Co., Ltd. \\
		Liquefied carbon dioxide & Showa Tansan Co., Ltd. \\
		\hline
		\multicolumn{2}{l} {* Only for solvogel synthesis. ** For other processes (e.g., rinse).} \\
	\end{tabular}
\end{table*}

Hydroxyl groups ($-OH$) on the surface of $SiO_2$ particles are likely to be charged and can easily react with other ions. Absorption of moisture into hydrophilic aerogels is a particularly serious problem. Hydroxyl groups are thus replaced with trimethylsiloxy groups [$-OSi(CH_3)_3$] by adding the hydrophobic reagent hexamethyldisilazane [$((CH3)_3Si)_2NH$] \cite{cite21}:
\[ 2(-OH) + ((CH_3)_3Si)_2NH \to 2(-OSi(CH_3)_3) + NH_3. \]
Water-resistant aerogels are obtained after supercritical drying (SCD) (see Fig. \ref{fig:fig1}).

Because the fine structure of silica networks is easily destroyed if solvogels are dried in air, they should be dried by the SCD method. When the solvent ethanol in solvogels is extracted from silica networks by natural evaporation, the path in the pressure-temperature phase diagram of ethanol intersects the boiling line. Because it is accompanied by significant ethanol-volume changes, the fine silica networks are lost. To avoid this, we change the solvent phase from liquid to gas by going around the critical point in the SCD method. The simplest way to extract ethanol is to go around the critical point of ethanol. The critical pressure and temperature of ethanol are 6.4 MPa and 243.1$^\circ$C, respectively. Although this approach is certainly available for at least $n < 1.02$, higher-refractive-index solvogels synthesized with DMF are broken in the ethanol SCD method. For this reason, solvogels in any refractive-index range are usually dried by the carbon dioxide SCD method. In this method, by using an autoclave, solvogels go around the critical point of carbon dioxide after the ethanol in the solvogel is replaced by liquefied carbon dioxide under high pressure. The critical pressure and temperature of carbon dioxide are 7.4 MPa and 31.1$^\circ$C, respectively. Using carbon dioxide is safe because it is nonflammable and has a low critical temperature.

\subsection{Solvogel synthesis}
\label{}
We begin by preparing two solutions A and B. Solution A is made by adding MS51 to the solvent and solution B is made by adding 28\% ammonia solution to distilled water. The starting materials for the aerogel production process are listed in Table \ref{table:table1}. Solutions A and B are quickly mixed in a polyethylene beaker at room temperature and stirred well for 30 s. The mixed solution is carefully poured into a polystyrene mold and immediately covered with a lid. To obtain more transparent aerogels, the amount of ammonia solution should be adjusted to complete the gel formation reaction in several minutes. Furthermore, depending on the solvent used, the solvogel is covered with a small amount of ethanol or methanol, and the mold is covered again with the lid to avoid evaporation of the solvent in the solvogel.

To strengthen the three-dimensional $SiO_2$ network, the synthesized solvogel in the mold is placed in an air-tight vessel and aged at room temperature for 1 week. In this aging period, the solvogel slightly shrinks and separates from the mold; thus, forming a solvogel tile.

\subsection{Hydrophobic treatment}
\label{}
After 1 week, the lid of the solvogel mold is removed and the air-tight vessel is filled with ethanol (99) so that the solvogel is immersed in ethanol. Next, the vessel is sealed again and the solvogel is left for 3 days to be impregnated with ethanol.

For the hydrophobic treatment, the solvogel is detached from the mold and transferred into a stainless steel punched tray in the same ethanol, and then the hydrophobic reagent is poured into the ethanol. The volume ratio of hydrophobic reagent to ethanol is 1:9. The solvogel is kept in the solution at room temperature for 3 days.

Ammonia generated by the hydrophobic reaction is extracted by significant quantities of ethanol, and the solvogel is then transferred into fresh ethanol. The ethanol is replaced by new ethanol every 2--3 days for a total of two times.

\subsection{Supercritical drying}
\label{}

\begin{figure}[t] 
\centering 
\includegraphics[width=0.50\textwidth,keepaspectratio]{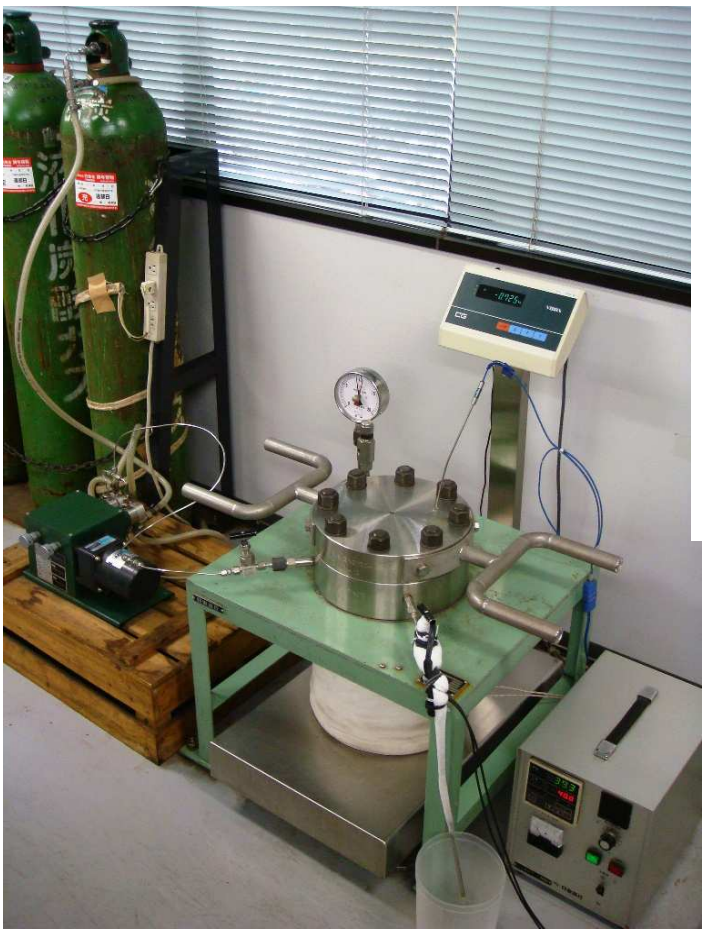}
\caption{Homebuilt supercritical extraction equipment ($CO_2$ autoclave) installed at Chiba University. The autoclave with a heater is mounted on an electronic balance. A pump is used to send liquefied carbon dioxide from a cylinder to the autoclave. Based on the temperature measured with a type-K thermocouple inserted into the autoclave, the output of the heater is adjusted by a controller. The autoclave has a capacity of 7.6 liters. Up to 10 solvogel tiles with dimensions of 11 $\times $ 11 $\times $ 2 cm$^3$ can be dried simultaneously. This size is sufficient for test production; for mass production of large aerogel tiles, we can use the large autoclave (140 liters) at the Mohri Oil Mill Co., Ltd.}
\label{fig:fig2}
\end{figure}

At this stage of the aerogel production, ethanol still remains in the voids between the silica networks of the solvogel. Ethanol can be extracted without cracking the solvogel by using the carbon dioxide SCD method. Supercritical extraction equipment (autoclave) was constructed for this purpose (see Fig. \ref{fig:fig2}).

\begin{figure}[t] 
\centering 
\includegraphics[width=0.50\textwidth,keepaspectratio]{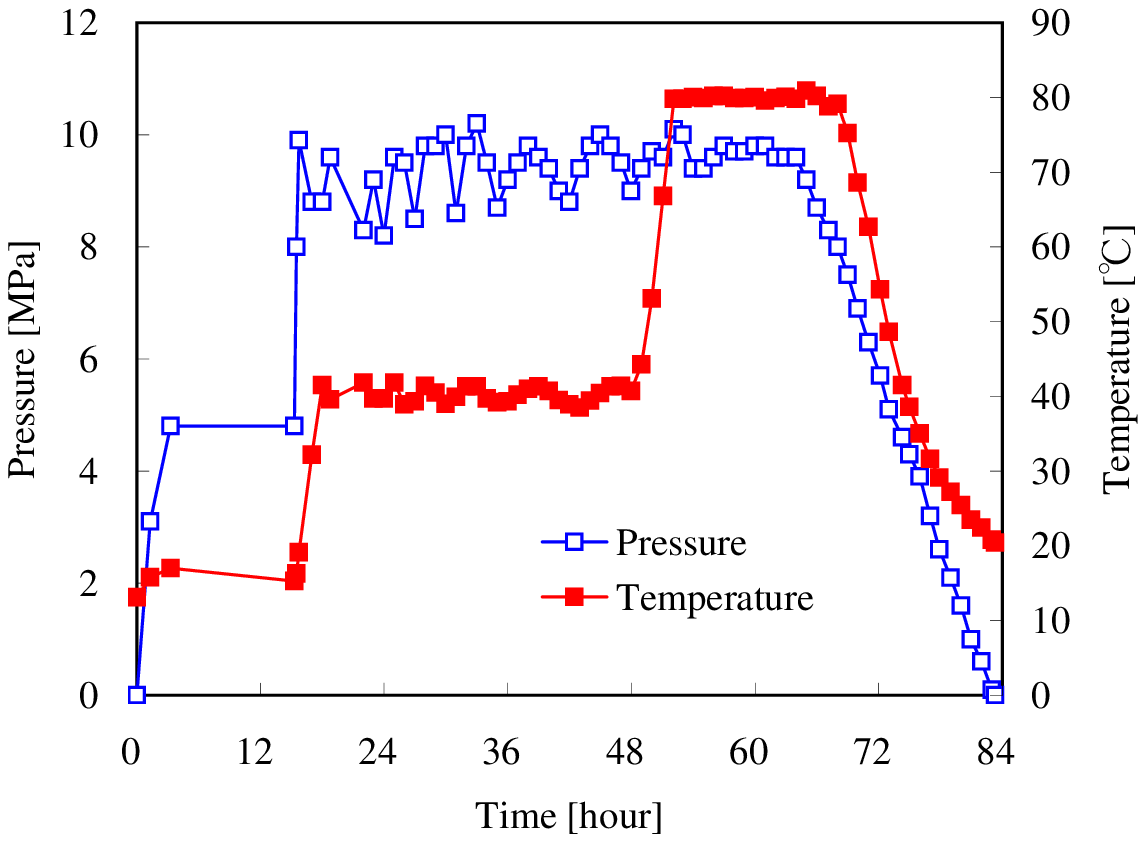}
\caption{History of internal pressure and temperature in the autoclave (run 143). The open and solid squares represent the changes in pressure (left axis) and temperature (right axis), respectively.}
\label{fig:fig3}
\end{figure}

After the solvogel is positioned in the autoclave, the autoclave is filled with fresh ethanol and sealed. Fig. \ref{fig:fig3} shows a history of the internal pressure and temperature in the autoclave. SCD begins with the free injection of liquefied carbon dioxide from a cylinder at room temperature. The injection stops when the autoclave pressure is equal to that of the cylinder, after which the autoclave is kept at the cylinder pressure ($\sim$ 6 MPa) for 1 night ($\sim$ 12 h). During this time, the liquefied carbon dioxide mixes well with the ethanol and penetrates the solvogel.

To increase the pressure as high as 8.0 MPa above the critical pressure of carbon dioxide (7.4 MPa), liquefied carbon dioxide is pumped into the autoclave. When the pump is stopped, the autoclave is heated at 10$^\circ $C/h to 40$^\circ $C above the critical temperature of carbon dioxide (31.1$^\circ $C). Note that the solvogel is below the subcritical condition because the autoclave is still rich in ethanol, whose critical temperature is 243.1$^\circ $C. The internal pressure in the autoclave increases with increasing temperature. When the pressure reaches 10.0 MPa, the extraction of the ethanol and carbon dioxide fluid mixture begins by opening an output valve to keep the pressure. Initially, almost all of the extracted fluid is ethanol.

When the temperature reaches 40$^\circ $C, injection by pumping liquefied carbon dioxide starts again. Extraction of the mixed fluid continues and the pressure is maintained in the range of 8.0 to 11.0 MPa. It takes 30 h from the beginning of the extraction before the extracted fluid is sufficiently rich in carbon dioxide. To extract all the ethanol from the solvogel, the temperature is raised to 80$^\circ $C at 10$^\circ $C/h. The conditions of 8.0$–-$11.0 MPa and 80$^\circ $C are maintained typically for at least 10 h. The concentration of ethanol vapor in the extracted fluid determines when the injection of liquefied carbon dioxide is finished. When an ethanol-inspection tube shows a concentration of 200 ppm or less, the pump is stopped.

The autoclave operation for the pressure- and temperature-reduction process is crucial for obtaining crack-free aerogels. Light, crack-free aerogels with $n \leq  1.03$ are relatively easy to make. However, a slow pressure-reduction rate is important for thick crack-free aerogels with $n \geq  1.04$. The pressure is first reduced below the critical pressure (7.4 MPa) at high temperature (80$^\circ $C). The rate of pressure reduction is 0.5$–-$1.0 MPa/h; in particular, 0.5 MPa/hour is recommended for solvogels thicker than 2 cm. Below the critical pressure, the temperature is gradually decreased together with pressure. When the ambient pressure and temperature are reached, the autoclave is still filled with carbon dioxide gas. After opening the autoclave, this gas is spontaneously replaced by air, which completes the aerogel production. In all, the carbon dioxide SCD process takes about 3.5 days. The shortest time to complete all the steps for aerogel production is 23 days.

\section{Optical measurements}
\label{}
For all aerogel tiles manufactured, the important optical parameters (i.e., refractive index and transmittance) were measured by hand, tile by tile, in the visible range. Owing to the recent progress in the RICH detectors by using an aerogel radiator, the need to focus on optical measurements is growing so that the Cherenkov angle from unscattered Cherenkov photons emitted by aerogels can be reconstructed.

\subsection{Refractive index}
\label{}
\begin{figure}[t] 
\centering 
\includegraphics[width=0.50\textwidth,keepaspectratio]{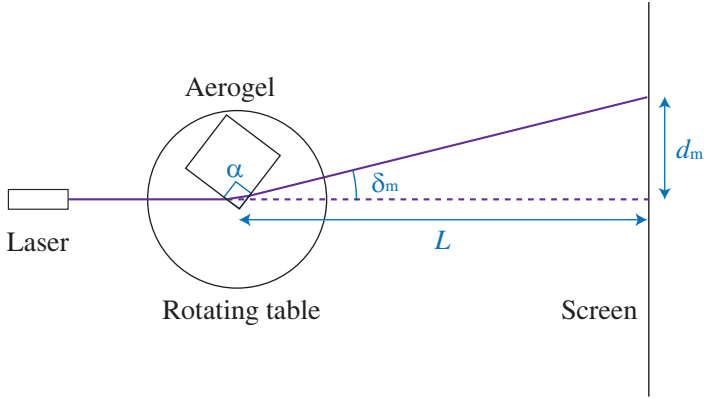}
\caption{Setup of refractive index measurement. The notation is explained in the text. The table can be manually rotated, and a grid sheet is fixed to the screen.}
\label{fig:fig4}
\end{figure}

Manufactured aerogels can be tagged by their refractive index at a wavelength ($\lambda $) of 405 nm, which can be measured by the laser Fraunhofer method with a blue-violet semiconductor laser. The measurement setup is shown in Fig. \ref{fig:fig4}. The Fraunhofer method gives the refractive index of aerogels relative to air and is based on the prism formula:
\[ \frac {n}{n_{air}} = \sin\left ( \frac {\alpha +\delta _m}{2}\right ) \left [ \frac{1}{\sin(\alpha /2) }\right ], \]
\[ \delta _m = \tan^{-1}(d_m/L), \]
where $n_{air} \sim  1.0003$ is the absolute refractive index of air, $\alpha  = \pi /2$ is the vertex angle of the aerogel, $\delta _m$ is the minimum angle of deviation, $d_m$ is the minimum displacement of the laser spot, and $L$ is the distance between the aerogel and the screen. The distance $L$ was approximately 1.8 m in our case. The averaged refractive index was determined by $d_m$ measured at four corners of each aerogel tile, where the laser penetration depth was 5 mm from the vertex of the aerogel. The laser-spot spread on the screen depends on the optical properties, including the aerogel refractive index. After passing through the aerogels, typical laser-spot diameters were 1, 6, and 16 mm (at equivalent resolution, $\Delta n =$ 0.0001, 0.0008, and 0.002) for $n \sim $ 1.01, 1.05, and 1.10, respectively.

\subsection{Transmittance (clarity)}
\label{}

\begin{figure}[t] 
\centering 
\includegraphics[width=0.50\textwidth,keepaspectratio]{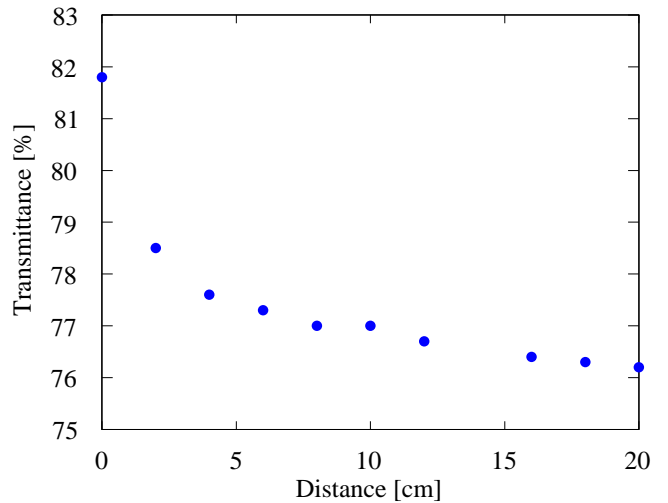}
\caption{Transmittance at 400 nm as a function of distance between the downstream surface of the aerogel (ID = PD156b, $n = 1.050$, and $t = 11.6$ mm) and the integrating sphere.}
\label{fig:fig5}
\end{figure}

The transmittance through 10 to 20 mm of aerogel was measured using the generalized spectrophotometer Hitachi U-4100 (or U-3210 prior to 2008). This system allowed us to measure transmittance from 200 to 800 nm. The measurement chamber consisted of a continuous light source, sample stage, and light-integrating sphere. The transmitted light was collected by the integrating sphere and introduced into a photomultiplier tube (PMT). The diameter of the entrance to the integrating sphere was 20 mm and, without aerogels, the spot of the light source spread to approximately 10 $\times$ 8 mm$^2$ at the entrance of the integrating sphere. Because the entrance of the integrating sphere was larger than the spot of the light source, a portion of the light scattered by the aerogel could enter the sphere. As a result, the transmittance for a given wavelength depends on the distance between the aerogel tile and the integrating sphere, as shown in Fig. \ref{fig:fig5}. To gather as effectively as possible only the unscattered light, we decided to place the downstream surface of the aerogel tile 10 cm in front of the entrance to the integrating sphere. This distance was measured for several samples and found to be reproducible.

\begin{figure}[t] 
\centering 
\includegraphics[width=0.50\textwidth,keepaspectratio]{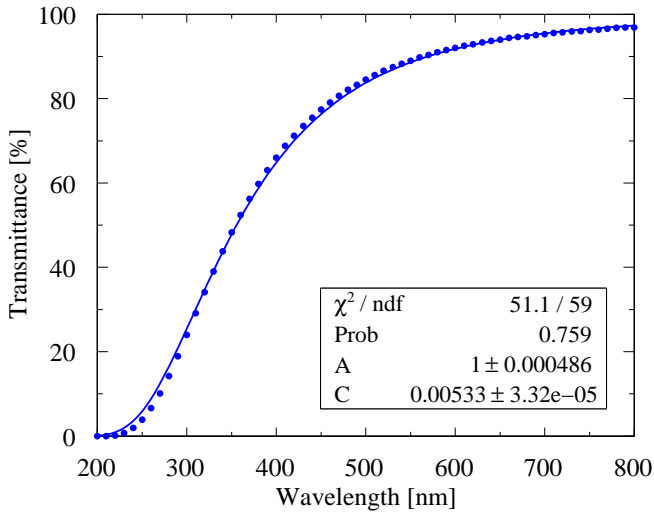}
\caption{Aerogel transmittance curve for aerogel ID = PDR21-2a, $n = 1.044$, and $t = 20.8$ mm. Circles show the transmittance measured every 10 nm by the spectrophotometer and the solid line shows the fit. The parameters obtained from the fitting with $T=A\exp(-Ct/\lambda ^4)$ are $A = 1$ and $C = 0.00533 \pm 0.00003$ $\mu $m$^4$/cm. The upper limit of the parameter $A$ was set to 1 in the fitting procedure. The corresponding transmission length was calculated to be 50 mm at $\lambda $ = 400 nm.}
\label{fig:fig6}
\end{figure}

Fig. \ref{fig:fig6} shows a resulting transmittance curve as a function of wavelength. The sample aerogel tile manufactured with DMF has a refractive index of $n = 1.044$ and a thickness of 20.8 mm. It is known that light transmission in aerogels is dominated by Rayleigh scattering:
\[ T(\lambda , t)=A\exp(-Ct/\lambda ^4), \]
where $T(\lambda , t)$ is the transmittance, $A$ and $C$ are parameters, and $t$ is the aerogel thickness. The parameter $C$ is called the "clarity coefficient" and is usually measured in units of $\mu $m$^4$/cm. The parameters obtained from the fitting are $A = 1$ and $C = 0.00533 \pm 0.00003$ $\mu $m$^4$/cm. The upper limit of the parameter $A$ was set to 1 in the fitting procedure.

\subsection{Transmission length}
\label{}
The transmission length $\Lambda _T(\lambda )$, which is defined as $\Lambda _T(\lambda ) = -t/{\rm ln}T(\lambda )$, is a useful parameter for comparing transparencies of aerogel tiles with differing refractive indices and thicknesses because, provided the surface effect can be neglected, $\Lambda _T$ is independent of thickness. We usually evaluate the transmission length at $\lambda $ = 400 nm, which is where typical photon detectors have peak quantum efficiency. The improvement of the transmission length for any given refractive index is our biggest goal.

\begin{figure}[t] 
\centering 
\includegraphics[width=0.50\textwidth,keepaspectratio]{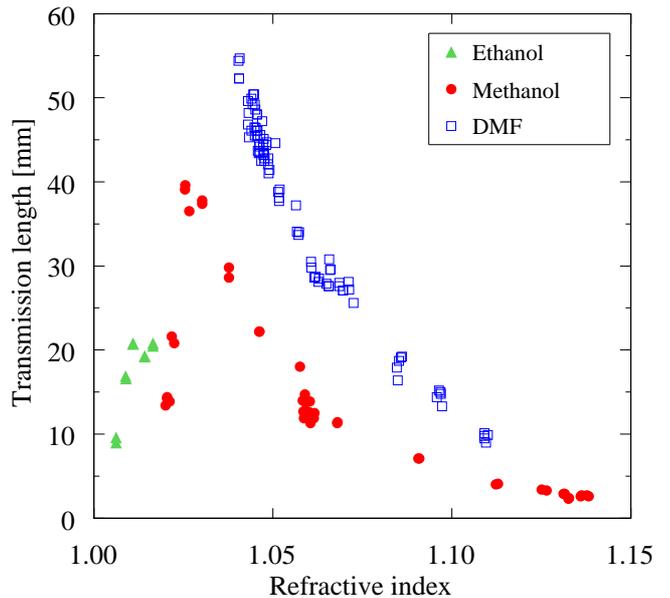}
\caption{Distribution of transmission length at $\lambda $ = 400 nm over a wide refractive-index range. The refractive index was measured at $\lambda $ = 405 nm. A total of 142 aerogel tiles are represented by separate dots: triangles, circles, and squares denote aerogels synthesized with ethanol, methanol, and DMF, respectively. The transmission length was calculated from transmittance measured with the spectrophotometer and from the measured thickness of the aerogels.}
\label{fig:fig7}
\end{figure}

Fig. \ref{fig:fig7} shows the distribution of transmission length in the range $n$ = 1.006$–-$1.138. The use of methanol as solvent provided a wide refractive-index range ($n > 1.020$). At $n = 1.025$ we obtained the most transparent aerogel tile ($\Lambda _T$ = 40 mm) for aerogels synthesized with methanol. For both $n < 1.025$ and $n > 1.025$, the transmission length decreased gradually. The introduction of DMF as solvent was tested in the range $n$ = 1.040$–-$1.110 and consistently provided aerogels more transparent than those synthesized with methanol. The longest transmission length of 55 mm was obtained at $n = 1.040$. However, the transmission length decreased rapidly to less than 45 and 30 mm at $n = 1.050$ and 1.060, respectively. Aerogels with $n < 1.020$ were synthesized using ethanol as solvent and resulted in a transmission length of 10 to 20 mm.

\section{Fine structure of silica aerogel}
\label{}
The transparency of aerogels depends on their nanostructure (i.e., their three-dimensional configuration) and the size of the primary $SiO_2$ particles and pores. The characteristic length scale of the nanostructure formed in the solvogel synthesis process was determined by scanning electron microscopy to be on the order of 10 nm. Experience shows that reducing the gelation time leads to more transparent aerogels. Aerogels synthesized using DMF are generally more transparent than those synthesized using methanol; a result that we attribute to the fact that DMF should form a finer $SiO_2$ structure. To verify this hypothesis, we performed SAXS experiments, which we describe next.

\subsection{Small-angle X-ray scattering}
\label{}
SAXS is the diffusive scattering produced by the contrast of the nanoscale electron density of the sample (i.e., Thomson scattering). A typical scattering angle is $2\theta $ = 0$^\circ $ to 5$^\circ $ with respect to the incident X-rays. For porous materials such as silica aerogel, information regarding the size and shape of the particles and pores may be obtained by SAXS because it is sensitive to structures 1 to 100 nm in size. Bragg's law applied to crystalline materials relates the scattering (or diffraction) angle to the lattice spacing $d$ of crystals:
\[ 2d\sin\theta = \lambda, \]
where $\lambda $ is the X-ray wavelength. In SAXS, the intensity $I$ of scattered X-rays is measured as a function of the scattering parameter $q = 4\pi \sin\theta /\lambda $. From Bragg's law and the definition of the scattering parameter, the structural spacing of crystals is given by $d = 2\pi /q$.

\subsection{SAXS measurements}
\label{}
We prepared approximately 1-mm-thin aerogel plates using methanol or DMF as solvent for SAXS measurements and also manufactured 1-cm-thick aerogels with a tile shape as a reference for optical measurements. Their refractive indices and transmission lengths were $n = 1.058$ and 18 mm, respectively, when methanol was used as solvent and $n = 1.061$ and 23 mm, respectively, when DMF was used as solvent.

The SAXS measurements were conducted at the Nishikawa Laboratory, Department of Chemistry, Chiba University on a NANO-Viewer (Rigaku) SAXS system, which consists of a monochromatized X-ray generator ($Cu$ K$\alpha $, $\lambda $ = 1.54 \AA), a focusing multilayer optic with three slits, and a detector. Except for the space where the sample holder was installed, the X-ray path was in vacuum ($< 100$ Pa) to avoid scattering of X-rays by air. X-rays scattered by the aerogel plates were detected with a two-dimensional imaging plate. The imaging plates were analyzed with an R-AXIS DS3C scanner (Rigaku). More details on the SAXS method are given in Ref. \cite{cite22}.

\subsection{SAXS results}
\label{}
\begin{figure}[t] 
\centering 
\includegraphics[width=0.50\textwidth,keepaspectratio]{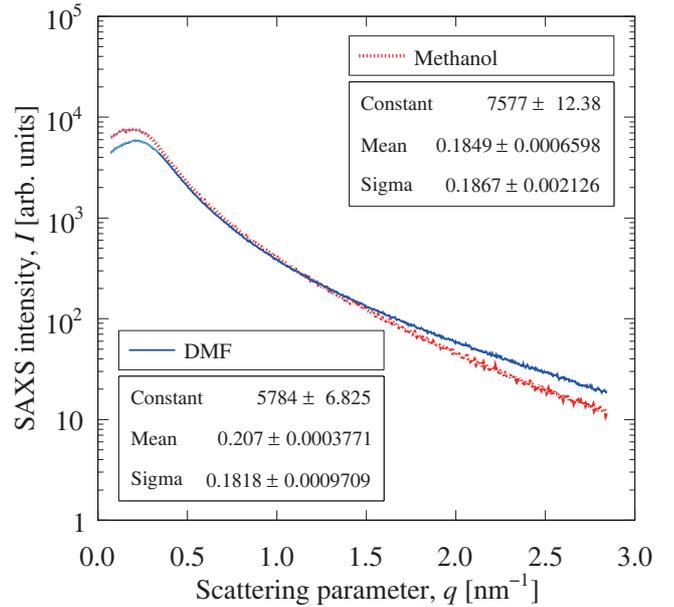}
\caption{Scattered X-ray intensity $I(q)$ as a function of scattering parameter $q$ for aerogels synthesized with methanol (dotted line) or DMF (solid line) as solvent. The SAXS profiles take into account the effects of the intensity fluctuation of the incident X-ray beam and the X-ray absorption of the aerogels. The intensities were fitted around the peaks by Gaussian functions, and the parameters obtained are shown on the graph.}
\label{fig:fig8}
\end{figure}

Fig. \ref{fig:fig8} shows the X-ray intensity as a function of the scattering parameter for aerogels synthesized with methanol or DMF. We conclude that the particle or pore shapes of the two aerogels were the same because the SAXS profiles are similar to each other. However, the two curves intersect. The scattering parameters taken at the peak intensities are $q = 0.1849 \pm  0.0007$ and $0.2070 \pm  0.0004$ nm$^{-1}$ for methanol and DMF, respectively, which means that the structural spacings were clearly different. Finer structural information appears at the higher range of scattering parameter. Although we should analyze the entire range of scattering parameter for more details \cite{cite23}, this result supports the conclusion that aerogels synthesized with DMF form a finer particle structure, making them more transparent than those synthesized with methanol.

\begin{figure}[t] 
\centering 
\includegraphics[width=0.50\textwidth,keepaspectratio]{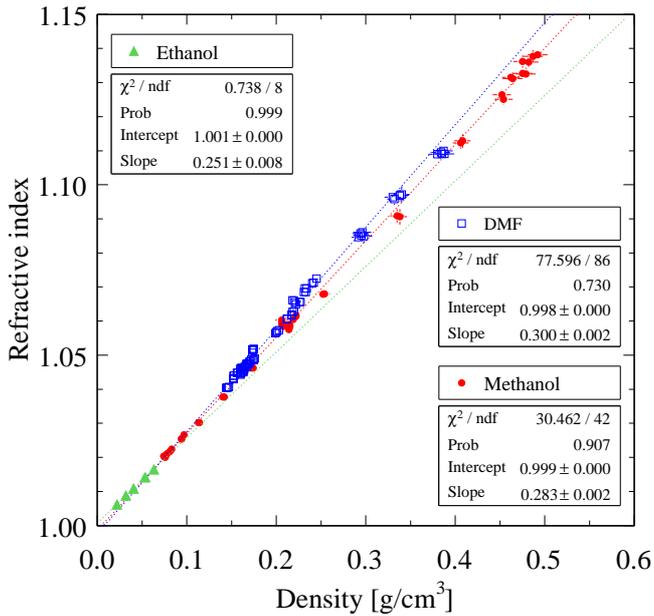}
\caption{Refractive index measured at $\lambda $ = 405 nm as a function of aerogel density. The aerogels, which are identical to those in Fig. \ref{fig:fig7}, were synthesized with ethanol, methanol, and DMF and are indicated by triangles, circles, and squares, respectively. The dotted lines represent best fits by linear functions to the three data series. The $k$ values are labeled as "Slope" in the legends.}
\label{fig:fig9}
\end{figure}

\begin{figure}[t] 
\centering 
\includegraphics[width=0.50\textwidth,keepaspectratio]{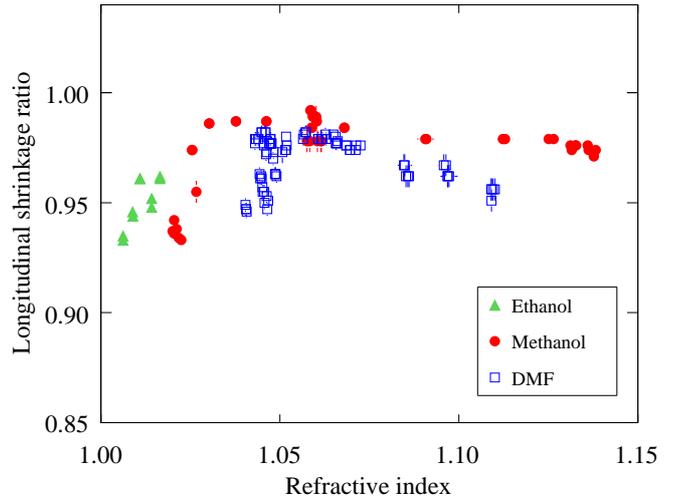}
\caption{Aerogel longitudinal shrinkage ratio as a function of refractive index (at 400 nm). Triangles, circles, and squares represent the aerogels synthesized with ethanol, methanol, and DMF, respectively.}
\label{fig:fig10}
\end{figure}

\section{Density measurement}
\label{}
When we use aerogels as a Cherenkov radiator, each aerogel tile is optically characterized by its refractive index, as described in Sec. 3. However, for aerogels used as cosmic dust collectors in planetary science, each aerogel tile is usually characterized by its bulk density. Acquisition of cosmic material is very important in planetary science. Micron-size dust particles traveling at hypervelocity (typically $\sim$ 10 km/s) in a low earth orbit (LEO) or in deep space can create impact craters on the aerogel surface and dig into the aerogel with impact tracks in the intact capture process. Thus, for this application, the density of aerogels is a barometer of the hypervelocity capture performance.

Although only aerogels with low refractive indices (below 1.016, which corresponds to a density $\rho \leq 0.06$ g/cm$^3$) are used as dust samplers, the density of all manufactured aerogels, including those with higher refractive indices, are always measured. From the mass, dimensions, and thickness, the density of aerogels can be gravimetrically determined. The mass is measured with an electronic balance to an accuracy of 10$^{-2}$ g. Our hydrophobic aerogels have no change in mass for at least ten years after production. The dimensions, including the thickness, are measured with a scale in increments of 0.25 mm along the four sides of the aerogel tile. The thickness measurement is also necessary to calculate transmission length. The form of an aerogel tile is fixed in the solvogel synthesis process, where the prepared solution coagulates with a meniscus on the surface of solvogels. Because of the meniscus geometry, aerogel tiles are approximately 0.5 mm thinner near the center. Transparent aerogels allow us to estimate average thicknesses through the sides. Note that the Panasonic products (e.g., SP-50) have no meniscus (thickness variation $\Delta t < 0.2$ mm) because they are surface treated in the solvogel synthesis process.

The refractive index is linearly related to the aerogel density by $n(\lambda ) = 1+k(\lambda )\rho $, where $k$ is a constant that depends on the wavelength of light and is given as 0.21 cm$^3$/g in the particle data booklet \cite{cite24}, which is consistent with the values given in Ref. \cite{cite25}. Our hydrophobic aerogels with added trimethylsiloxy groups show larger $k$ to some extent. Fig. \ref{fig:fig9} shows the relationship between refractive index and density for aerogels synthesized with ethanol, methanol, and DMF. We found different values for $k$ depending on the solvent used: at $\lambda $ = 405 nm, $k$ = 0.251, 0.283, and 0.300 cm$^3$/g for ethanol, methanol, and DMF, respectively. This result reflects the nanostructural differences between aerogels synthesized with different solvents. Although the difference in $k$ between methanol and DMF is small, the refractive index of aerogels synthesized with DMF is always larger than that of aerogels synthesized with methanol (at the same density).

The measurement of dimensions is advantageous for computing the shrinkage ratio of aerogel tiles. From its evaluation, we can understand that fluctuations in the refractive index between several aerogel tiles or between production batches are attributed in large part to fluctuations in the shrinkage ratio. The longitudinal shrinkage ratio as a function of refractive index is shown in Fig. \ref{fig:fig10}. This ratio is defined as $l/l_0$, where $l$ is the length of the sides of a manufactured final aerogel and $l_0$ is the size of the molds for the solvogel synthesis. In Fig. \ref{fig:fig10}, we see two groups of data near $n \sim 1.045$ for aerogels synthesized with DMF. This result is a reflection of the dependence of the shrinking ratio on aerogel size; that is, in this case, small aerogels (52 cm$^3$) shrank more than large aerogels (183 cm$^3$). When we adjust the refractive index in aerogel preparation, it is important to take into account the shrinkage.

\section{Conclusion}
\label{}
We characterized our original (and conventional) method of producing hydrophobic silica aerogel. Since the early 2000s, several thousand aerogel tiles have been manufactured at Chiba University and Panasonic Electric Works under the Belle upgrade program and for other purposes. We examined all the aerogels produced, and 142 samples were selected and described in detail in terms of their basic properties. The introduction of DMF as solvent in the solvogel synthesis process resulted in improvements in aerogel transparency in the high-refractive-index range. SAXS measurements revealed that finer aerogel structure can be formed by using DMF as solvent.

\section*{Acknowledgments}
\label{}
We are grateful to the members of Particle Physics Laboratory of Chiba University for their assistance in aerogel production. We are also grateful to Dr. Y. Hatakeyama of Chiba University for his full support with the SAXS experiments. This work was partially supported by a Grant-in-Aid for JSPS Fellows (No. 07J02691 for M.T.) and a Grant-in-Aid for Scientific Research (C) (No. 17540284 and 19540317 for I.A.) from the Japan Society for the Promotion of Science (JSPS) and a Grant-in-Aid for Scientific Research on Innovative Areas (No. 21105005) from the Ministry of Education, Culture, Sports, Science and Technology (MEXT). This publication was in part supported by the National Institute for Fusion Science (NIFS) in the National Institute of Natural Sciences (NINS) of Japan.








\end{document}